# What happens when the geomagnetic field reverses?

J.F. Lemaire* and S.F. Singer **

* BISA, Brussels, *joseph.lemaire@aeronomie.be* ;
** University of Virginia / SEPP, Arlington, USA, *singer@sepp.org*

**ABSTRACT.** During geomagnetic field reversals the radiation belt high-energy proton populations become depleted. Their energy spectra become softer, with the trapped particles of highest energies being lost first, and eventually recovering after a field reversal. The radiation belts rebuild in a dynamical way with the energy spectra flattening on the average during the course of many millennia, but without ever reaching complete steady state equilibrium between successive geomagnetic storm events determined by southward turnings of the IMF orientation. Considering that the entry of galactic cosmic rays and the solar energetic particles with energies above a given threshold are strongly controlled by the intensity of the northward component of the interplanetary magnetic field, we speculate that at earlier epochs when the geomagnetic dipole was reversed, the entry of these energetic particles into the geomagnetic field was facilitated when the interplanetary magnetic field was directed northward. Unlike in other complementary work where intensive numerical simulations have been used, our demonstration is based on a simple analytical extension of Störmer's theory.  The access of GCR and SEP beyond geomagnetic cut-off latitudes is enhanced during epochs when the Earth's magnetic dipole is reduced, as already demonstrated earlier.

## 1. Introduction

Undoubtely, geomagnetic field polarity reversals have had drastic effects on the inner radiation belts, as well on the access of Galactic Cosmic Rays (GCR) and Solar Energetic Particles (SEP) into the magnetosphere. Already Uffen [1963] speculated that *"During those intervals the trapped corpuscular radiation may have been spilled on the Earth, ..."* . Solar Proton Events produce ozone depletions (e.g. Jackman et al., [2005]), and the access of Galactic Cosmic Rays may possibly control the cloud coverage, and thus influence the Earth's climate [Svensmark and Friis-Christensen, 1997; Kirkby et al., 2011].  The historical development and perspectives of magnetic polarity transitions on these geophysical phenomena, as well on the biosphere have been updated by Glassmeier and Vogt [2010] (see also the book of Glassmeier and Soffel [2009]).

Directly related to the present study, Stadelmann et al. [2010] presented results based on thousands of trajectories calculations of GCR and SEP in a potential B-field model of paleomagnetospheres where the dipole and quadrupole moments have given prescribed values. Like Smart et al. [2000] they determined the geomagnetic cut-off latitudes and "impact areas" which quantify the percentage of the Earth's surface that is accessible to charged particles of a given energy, from all incident directions and from all locations on a spheric shell in outer space. These detailed calculations as well as those of Vogt and Glassmeier [2000], Vogt et al. [2004, 2007], Zieger et al. [2004, 2006] are interesting and suitable for non-symmetrical magnetic field configurations. Emphasis in these earlier studies was mainly on the impact of GCR and SEP on the atmosphere. The main objectives of this study relate to the energy spectra of trapped radiation belt particles. The key role of the changes of orientation of the IMF during  geomagnetic field reversals will be pointed out.

In the following we use a less demanding analytical approach based on Störmer's theory which was originally developed for a dipole magnetic distribution [Störmer, 1907, 1913, 1955]. This theory has been extended by Lemaire [2003] by adding a uniform interplanetary magnetic field ($F$) to the magnetic dipole ($M$). As a consequence of its zonal symmetry $\phi$, the azimuthal/longitudinal coordinate is a dummy variable in the expression of the Hamiltonian of charged particles moving in this B-field configuration.  Therefore, topologically allowed and forbidden zones can be defined for incoming and trapped particles of prescribed kinetic energy or magnetic rigidity. This property enables us to determine rather easily - analytically, without intensive numerical integration of particle trajectories - the geomagnetic cut-off surface, as well as the guiding center field line of trapped radiation belt particles (i.e. their "Thalweg") for different orientations of the IMF (**F**) and for different values of the Earth's dipole moment (**M**).

To set the stage, Störmer theory will first be briefly reviewed. It will then be shown how the Thalweg of a trapped radiation belt particle of given energy breaks open by interconnecting to the IMF when the latter turns southward and becomes sufficiently negative. Conversely, interplanetary ions will then be able to enter in the inner part of the magnetosphere, and possibly even impact the atmosphere.

Our analytical approach shows also that the energy for which trapped radiation belt particles can spill out of the trapping zones is gradually reduced when $M$ tends to zero during a geomagnetic field polarity transition.

## 2. Brief review of Störmer's theory

*Coordinate systems.* Störmer [1907] found that trajectories of particles in a dipole B-field are confined within regions corresponding to *inner and outer allowed zones* subsequently named after him. The allowed zones are separated by a *forbidden zone* whose shape is a function of the constants of motion of charged particles: the energy, and the generalized or canonical angular momentum. These zones are well described by Störmer [1955], where he uses a coordinate system whose unit length is proportional to $Ze$, the electric charge of the particle, and inversely proportional to the modulus of generalized momentum, $\mathbf{p} = m \mathbf{v}$.

Another, but less well known, mathematical formulation of Störmer's theory was proposed many years later by Dragt [1965]. The latter is simpler and leads to more intuitive graphical representations of Störmer's allowed and forbidden zones. Dragt introduced the dimensionless time, $t$, and polar coordinates ($\rho, z, \phi$), whose units are determined respectively by

$$t_u = |m (Ze M)^2 / p_o^3| \quad (1) \quad \text{and} \quad r_u = |Ze M / p_o| \quad (2)$$

where $m$ is the mass of the particle, and $p_o$ is the (constant) azimuthal component of the generalized/canonical momentum ($p_\phi$). The latter is a constant of motion due to the zonal/axial symmetry of the dipole magnetic field distribution. For trapped radiation belt particles the value of $r_u$ is larger than $R_E$, the radius of the Earth ($r_E = R_E/r_u < 1$). The relationship between $r_u$ and $t_u$ was illustrated by Lemaire [2003] in his Figure A4.

*The Störmer potential.* Using Dragt's coordinate system the distribution of the Störmer potential, $V(\rho, z)$, is illustrated in Figure 1 by iso-contours ranging from $V_0 = 0$ to $\infty$ in a meridional plane. This 2D representation is independent of $\phi$, the longitude of the meridional plane. A corresponding 3D landscape representation of the function $V(\rho, z)$ was illustrated in Figure 5a by Lemaire [2003].

In such 2D and 3D plots for particles for which $Ze$ and $p_o$ have the same algebraic signs, there is a saddle point at $z=0$ and $\rho=2$, where two iso-contours cross each other. At this saddle point the equatorial potential distribution $V(\rho, 0)$ has a maximum value, $V(2,0) = 1/32$. It can be seen from Figure 1 that beyond this saddle point Störmer's potential tends to zero when $r = (\rho^2 + z^2)^{1/2} \to \infty$. In the opposite direction $V(\rho, z)$ forms a deep "valley" whose steep earthward wall corresponds to the "geomagnetic cut-off surface".

The bottom of this "valley" indicated by the dotted line in Figure 1 is where $V = 0$. It corresponds to the minimum of Störmer's potential which is called the Thalweg (path in the valley). Indeed, the function $V(\rho, z)$ can be compared to a geopotential surface whose "height" is proportional to $V(\rho, z)$. Any particle inside this valley is trapped within Störmer's inner allowed zone when its dimensionless kinetic energy is smaller than *1/32*. The projection of the trajectory of trapped particles in a meridional plane is then strictly confined between two of the dashed lines shown in Figure 1, which constitute the borders of the inner Störmer zone (e.g. *V = 1/1000*).

*The Thalweg and the guiding center field line.* The Thalweg at the very bottom of this "valley" coincides with the dipole magnetic field line along which trapped particles of small kinetic energy (KE < 1/32) spiral, bounce and drift indefinitely. In Dragt's coordinate system the Thalweg corresponds to the dipole magnetic field line traversing the equatorial plane at $\rho = r_T = 1$.

The Thalweg corresponds to the guiding center field line in Alfvén's adiabatic perturbation theory [Alfvén, 1940, 1950]. The equatorial crossing points of Thalwegs correspond to the central points of the bounce motion of trapped particles, and are part of what Vogt and Glassmeier [2000] label the "trapping center surface" in their paleomagnetosphere models.

*Access of high energy interplanetary particles into the inner geomagnetic field.* Interplanetary particles from infinity will move around the dipole either eastward or westward depending on the sign of their electric charge and of angular momentum. When their dimensionless kinetic energy (KE) is smaller than 1/32 their trajectories remain confined in the outer allowed zone, beyond a distant iso-contours of Figure1 (e.g. the dashed curve labeled *V = 1/53*).

Only interplanetary particles for which KE > 1/32 can transit over the saddle point region: i.e. over what happens to be a forbidden zone for particles of lower energies (KE < 1/32). Consequently, unless their kinetic energy exceeds the dimensionless threshold of 1/32, GCR or SEP are unable to penetrate deep into the geomagnetic field and populate the inner allowed Störmer zone.

*Sources of radiation belts particles.* Van Allen and Singer [1950] made good use of Störmer's theory to deduce the energy spectrum of primary cosmic rays by measuring their flux in rocket experiments at different geomagnetic latitudes. After the discovery of the radiation belts by Van Allen et al. [1958], Singer [1958a & b; 1959a & b] proposed cosmic-ray albedo neutrons as a source for the corpuscular radiation populating Störmer's inner allowed zone. Note that this same injection mechanism was also proposed independently by Vernov et al. [1959] as a source of the trapped energetic particles detected both by the Sputnik and Explorer spacecraft. The source and loss processes for radiation belt particles were comprehensively reviewed by Walt [1996] and by Singer and Lemaire [2009].

*The renaissance and fall of Störmer's theory.* Thus Störmer's theory became again popular in 1958, after the discovery of the radiation belts. It looked to be a very promising tool until 1961, when McIlwain's invariant geomagnetic coordinate system proved to be very useful to map fluxes of energetic electrons and protons trapped in the geomagnetic field [McIlwain, 1961, 1966]. This gave new life to Alfvén's "first order guiding center theory" also promoted by Northrop [1963]. Unfortunately this contributed to the fall/decline of Störmer's theory from theoretical space physics. Nevertheless, at least for a few years, Störmer's theory had been successfully used in studies of trajectories of 0.1-10 MeV electrons, and 0.5-300 MeV protons trapped within the radiation belts. See also S. Akasofu [EOS 84, 22 July 2003] for a critical discussion.

### 3. An extension of Störmer's theory

*Magnetic field line distribution.* As has been recalled above the cylindrical symmetry of this B-field distribution allows to reduce from 3 to 2 the number of independent variables to describe the trajectory of a charged particle. Indeed, once $\rho(t)$ and $z(t)$ are calculated in a meridional plane as a function of time *t*, the azimuthal velocity, $v_\phi = \rho \, d\phi/dt$ is also determined. By integrating the azimuthal component of the equation of motion, $\phi(t)$ can then be directly calculated.

Less than a decade ago, it was found that the same procedure can apply in the cases of more general B-field distributions or models: e.g. when a uniform interplanetary magnetic field (***F***) is superimposed on the dipole magnetic field (***B**_d*), as for instance in Dungey's [1961] magnetic field model. Indeed, when this interplanetary magnetic field (IMF) is either parallel or anti-parallel to the dipole moment, ***M***, the cylindrical/zonal symmetry of the B-field and of the Hamiltonian is preserved. Under such circumstances $p_\phi$, the azimuthal component of the generalized momentum, is again a con-stant of motion, $p_\phi = p_o$, and Lemaire [2003] pointed out that an extended Störmer potential, $V(\rho, z)$, can again be defined (cnf. http://arxiv.org/abs/1207.5160).

Let $b$ be a normalized value of $F$, the northward component of the IMF, namely,

$$b = F r_u^3 / 2 M \qquad (3)$$

where $M/r_u^3$ is the magnetic field intensity in the equatorial plane at $r = r_u$. The relationship between $b$ and $F$ has been illustrated in Figure A5 by Lemaire [2003] for different values of $r_u$ when the Earth reference dipole magnetic moment is $M_E = 8.06 \; 10^{15}$ Tesla m$^3$.

For a fixed value of $b$ the dimensionless equation of magnetic field lines is given by

$$r = k \cos^2(\lambda) \, [1 - b \, r^3] \qquad (4)$$

The whole family of magnetic field lines is generated by varying $k$ from 0 to ∞.

For $b > 0$ (northward IMF) all magnetic field lines that are traversing the Earth surface are "closed". They cross then the equatorial plane at $r = \rho < k$. The equatorial distance of the Thalweg $\rho_T$ is a solution of the algebraic equation: $\rho_T + b \, \rho_T^3 - 1 = 0$.

For southward IMF orientation $b < 0$, the total magnetic field, $\mathbf{B}_d + \mathbf{F}$, vanishes along the equatorial circumference of radius $\rho_x = 1 / (-2 b)^{1/3}$. This circumference corresponds to the X-line (or neutral line) of Dungey's magnetospheric model.

*The equatorial cross-section of the extended Störmer potential.* When the interplanetary magnetic field ($\mathbf{F}$) is parallel or anti-parallel to the dipole magnetic moment ($\mathbf{M}$) an axially symmetric magnetic potential, $V(\rho,z)$, can still be defined as in Störmer's theory [Lemaire, 2003]. Figure 2 shows the equatorial cross-section of $V(\rho, 0)$ for $b = 0$, $-0.03$, $-0.05$, $+0.05$ and $+0.03$.

The solid curve in Figure 2 corresponds to Störmer's equatorial potential distribution for a dipole, i.e. when $F = b = 0$. The four other curves correspond to non-zero values of the IMF; the dashed lines are for southward IMFs ($b < 0$), the dotted lines are for northward ones ($b > 0$).

The maximum of $V(\rho, 0)$ is at an equatorial distance $\rho_{max}$ which is solution of the equation: $\rho_{max} - b \, \rho_{max}^3 - 2 = 0$. When $b = 0$, Störmer's values are recovered: $\rho_{max} = 2$, $\rho_T = 1$, and $\rho_X = \infty$.

It can be seen that when $b$ becomes more negative, the maximum "height" of the potential decreases below the value $V_P = 1/32$. Conversely, when the IMF is northward and assumes larger values, $V(\rho, 0)$ increases. This implies that larger kinetic energies are required for charged particles to spill out the inner trapping zone as well as to enter into it. Thus Figure 2 illustrates how the IMF orientation and intensity controls the energy threshold of charged particles that can exit the radiation belts, and enter the magnetosphere or a paleomagnetosphere whose magnetic moment would be reduced.

*Critical Thalweg field lines.* In Störmer's theory, the Thalwegs are always "closed". In Lemaire's extended theory the Thalweg is not always "closed": it can become "open" when $b < b_T = - 4/(3)^3 = - 0.148148$. The Thalweg is then formed of two separate magnetic field line segments extending from the Earth surface to infinity in both hemispheres.

The threshold value $b_T = - 0.148148$ corresponds to a critical value for which $\rho_T = \rho_X$. The dashed line in Figure A5 of Lemaire [2003] indicates that this critical value of $b_T$ corresponds to $F = -42.75$ nT, $-9.2$ nT, and $-3.4$ nT, respectively for $r_u = 6R_E$, $10R_E$ and $14R_E$. The same critical value $b_T$ can as well be obtained by reducing $M$, the magnetic moment of a paleomagnetosphere by factors of $42.75$, $9.2$ and $3.4$, respectively with respect of $M_E = 8.06 \; 10^{15}$ Tesla m$^3$.

For $b < b_T$ the iso-contours of the extended Störmer potential are illustrated in Figure 3, using the same representation as in Figure 1. The two "open" branches of the Thalweg are then located in the middle of the dark-red cusp regions (valleys); these semi-infinite field line segments extend in both hemispheres from the Earth surface to infinity ($z = \pm\infty$). They are not drawn in Figure 3.

Therefore, when $b < -0.148148$, the Thalweg magnetic field lines are interconnected to the interplanetary magnetic lines. Whether these field line segments are interconnected or reconnected is a matter of terminology or semantics beyond our preoccupation in the present study!

*Geomagnetic cut-off.* The geomagnetic cut-off surface is the locus of points where the meridional component of the velocities becomes equal to zero: i.e. where $v_\rho^2 + v_z^2 = 0$, and $\mathbf{v} \cdot \mathbf{B} = 0$. When a GCR particle reaches the geomagnetic cut-off surface its velocity is normal to the meridian plane, and $v_\phi = \rho \, d\phi/dt = v$. Note that in Alfvén's "*first order guiding center theory or approximation*" the geomagnetic cut-off was not explicitly defined. It can, however, be assimilated to the "mirror points" where $v_\phi = v$ and $\mathbf{v} \cdot \mathbf{B} = 0$.

In Figures 2 and 3, the geomagnetic cut-off is located along the inner "slope" of the "valley". Its radial distance ($r_G$) is always smaller than $\rho_T$. The latitude ($\lambda_G$) where it penetrates the Earth's atmosphere can be approximated by $\lambda_G \cong arcos(r_E)^{1/2}$ for radiation belts particles for which $r_E^2/4\gamma_I^2 \ll 1$ and when $-b\, r_E^3 \ll 1$: cnf. eq. (19) in Lemaire [2003; http://arxiv.org/abs/1207.5160].

It can be verified that $\lambda_G$ and $r_G$ are not very sensitive functions of the interplanetary magnetic field component $F$ or $b$. However, these quantities characterizing the geomagnetic cut-off positions depend significantly of the value of $M$, as well as on the magnetic rigidity which is equal to $p_o/Ze$ as verified by Smart and Shea [1967].

These results are well supported by GCR observations indicating that the geomagnetic cut-off latitude decreases as a function of the magnetic rigidity, but that their fluxes are almost not affected by the orientation of the IMF. These results from Lemaire's extended Störmer theory confirm also that $\lambda_G$ and $r_G$ decrease when $M$, the dipole moment of paleomagnetospheres is reduced, as also shown from trajectory calculations by Vogt et al. [2007].

## 4. What happens when the Earth's magnetic moment reverses?

It has been considered above that the geomagnetic field distribution can be modeled (approximated) by a stationary superposition of (i) a dipole whose magnetic moment $M$ is directed anti-parallel to the $oz$-axis, and (ii) a uniform IMF whose intensity $F$ is either parallel or anti-parallel to this axis. Of course, other cylindrically symmetric/zonal, magnetic field distributions can be considered for which $F$ is an analytical function of $\rho$ and $z$, but not of $\phi$. The Dst-field generated by a symmetrical ring current is just the next simplest example. Special paleomagnetospheres where the quadrupoles and multipole components have a zonal symmetry, as in studies of Vogt and Glassmeier [2000] and Vogt et al. [2004, 2007], are other B-field distributions for which a generalized Störmer potential, $V(\rho, z)$ could be defined, in principle.

Such generalization of Störmer's potential would clearly improve the picture but would complicate the mathematics at the expense of analytical simplicity. Thus, for the sake of simplicity and illustration we considered a uniform IMF as an ideal B-field model to outline what might happen when $M$ experiences a secular variation and possibly reverses.

Let us now slowly decrease the value of $M$, but without changing the value of $F$. According to equation (2), the unit length, $r_u = |Ze\, M/p_o|$ decreases with $M$ for a particle of given charge ($Ze$) and a constant canonical angular momentum ($p_o$).

When $M$ decreases the modulus of $b = F\, r_u^3 / (2\, M)$ increases, and the extended potential displayed in Figure 2 will either grow or drop below the solid curve, depending on the initial orientation of the IMF. Remind that, according to equation (3), $b$ is half the ratio between the IMF intensity and that of the Earth dipole at $\rho = 1$ in the equatorial plane.

Now when the IMF changes from a northward to a southward orientation, $V(\rho, z)$ switches rapidly from one of (upper) dotted curves for which $b <0$, to a (lower) dashed curve for which $b <0$, ...and vice-et-versa when the IMF turns back northward. Therefore, at each southward turning of the IMF a certain fraction of trapped radiation belt particles can escape from the geomagnetic field provided their kinetic energy exceeds the maximum of $V(\rho, z)$.

The smaller the value of *M,* the lower is the energy threshold of radiation belt particles which can escape from the paleomagnetosphere. Extremely negative values of *b* are more frequently obtained when *M* is much smaller than its present day value.

For sufficient small value of *M*, the value of *b* can become more frequently smaller than the critical value $b_T = -0.148148$ when the IMF fluctuates as usual. The Thalwegs of trapped radiation belt particles will then break open more often into two separate branches interconnecting the inner region of the geomagnetic field with the IMF. When such extreme situations occur, trapped radiation belt particles of lower energies and rigidities can escape along these open Thalweg segments. They are lost to interplanetary space more frequently. Note that on such IMF polarity changes solar wind particles might also more easily rain into the inner magnetosphere over the polar caps.

Not to say, of course, that our simplified scenario deliberately ignores additional effects that solar wind induced electric fields and time dependent magnetic fields will impose in the magnetopause and magnetosheath regions on escaping magnetospheric particles of a few tens of keV or less.

Anyway all this implies that the trapping lifetime and the residual flux of energetic radiation belt particles must be dramatically perturbed and reduced at epochs of geomagnetic polarity transitions, … at least much more than nowadays. Similar conclusions had already been obtained by Zieger et al. [2004], Vogt et al. [2007] from their detailed trajectory calculations in paleomagnetospheric models. See also the review by Glassmeier and Vogt [2010].

Note that after a geomagnetic field reversal, when *M* has eventually recovered the same magnitude, but of opposite sign, the unit length ($r_u$), the geomagnetic cut-off latitude and Thalweg latitude ($\lambda_G$ and $\lambda_T$) recover the same values as before the GF polarity transition. Note that in the caption of Figure 2 the orientation of the IMF must be changed, however. Interplanetary particles will then have easier access to the inner magnetosphere for northward IMF orientation, instead of the southward one as happens to be the case today. Furthermore, storm time ring currents will then build up when the IMF is turning northward instead of southward as nowadays.

## 5. Energy spectrum of radiation belt particles during geomagnetic reversal

Based on this scenario, we can now infer what happens when the dipole field weakens and reverses as assumed above. Trapped protons are characterized by their kinetic energy and their rigidity, $B r_L$ which is equal to $p_o / Ze$. For GCR or radiation belt particle of given rigidity the gyro-radius, $r_L$, must increase when $M(t)$ and $B(r, t)$ decrease. This causes the proton to intersect the higher density portions of the atmosphere at lower altitudes and lower latitudes. As a matter of consequence, this causes an enhanced loss of energy and momentum of the trapped particles, and constitutes therefore an additional mechanism which limits the lifetimes of radiation belt particles, as first pointed out and modeled by Griem and Singer [1955]. See also the review by Singer and Lenchek [1962a].

On the outer edge of the proton belt $r_L$ increases until it becomes of the order of the Earth radius – at which point adiabatic invariance breaks down [Singer 1959a & b]. According to Lemaire [1962, 1963] the equatorial plane is indeed the most efficient place to violate the conservation of the first and second adiabatic invariants of particles of a given electric charge, mass, and kinetic energy. This implies that charged particles with equatorial pitch angles close to 90° are more prone to become untrapped than those mirroring at higher latitudes [Lemaire, 1962, 1963]. Note in Figures 1 and 3 that it is precisely close to the equatorial plane where the inner and outer allowed zones of Störmer are interconnected, that particles are able to spill out of the inner magnetosphere, nowhere else.

Of course, Alfvén's perturbation theory can also break down as a consequence of rapid time variations of the B-field, for instance due to the effect of hydromagnetic waves as originally discussed by Dragt [1961]. Resonant wave-particle interactions with magnetospheric ULF and whistler waves have such non-adiabatic effects on the trapped radiation belt particles as continuously proposed (see for instance Horne et al. [2005] and references therein).

Whether these non-adiabatic processes are able to fully account for the rapid "non-adiabatic acceleration and losses" observed with Explorer 15 and reported by McIlwain [1963, 1996], remains to be proven and will demand additional independent investigations.

Anyway, from the discussion above, we conclude that during the geomagnetic field reversal the highest energy protons will be lost first. The energy spectrum of trapped proton gradually steepens, as pointed in a review by Glassmeier and Vogt [2010].

As demonstrated by radiation belt measurements of McIlwain [1963, 1996] and others, the radiation belt particle population never reaches a stationary equilibrium spectrum. This is due to the endless variations of the IMF, and the continuous pitch-angle scattering of trapped particles by their non-resonant wave-particle interactions with magnetospheric ULF and whistler waves. The smaller the paleomagnetic dipole moment, (i) the larger will be the variations of particle fluxes remaining trapped, (ii) the lower will be their maximum energy threshold, and (iii) the softer will be their average energy spectrum.

When the geomagnetic field eventually recovers but with a reversed polarity, the reverse scenario takes place: average energy spectrum gradually flattens and becomes harder, maximum energy threshold for which radiation belt particles remain trapped is gradually increasing.

We hope that the present reconsideration of Störmer's theory will be useful from a historical perspective and will inspire younger researchers who may not have been exposed to Störmer's seminal theory. We expect that our extension of this theory by adding a simple uniform IMF to Störmer's dipole B-field model, has given some insight into the not yet fully exploited potentialities of this approach in investigations on the role (i) of southward/northward turning of the IMF, and (ii) of geomagnetic field reversals on the energy spectrum of radiation belt particles, and on the geomagnetic cut-off of GCRs, or (iii) on the formation of ring current, and (iv) on the generation of McIlwain's [1974] substorm injection boundaries in today's magnetosphere as well as in paleomagnetospheres.

**Acknowledgments**  We wish to thank V. Pierrard, M. Echim and D. Summers for editing various versions of this manuscript. We appreciate also the useful comments and suggestions of both referees.

### List of references

**Caption of Figures**

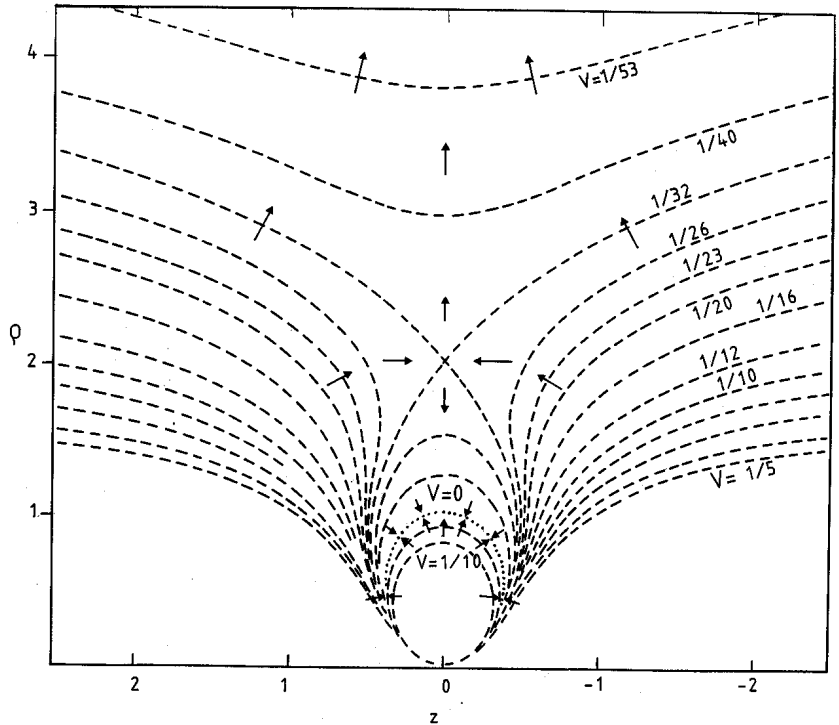

Figure 1. *Störmer's dimensionless potential, V(ρ, z)*. The dashed lines are the isocontours of V(ρ,z). They determine the frontiers between the Störmer's allowed and forbidden zones for V ranging from V = 0 to 1/5. The dotted line is the Thalweg which coincides with the dipole magnetic field line around which trapped particle spiral and oscillate between conjugate mirror points. The arrows indicate the slopes of V(ρ, z). The potential has maximum (V = 1/32) at z = 0 and $\rho_{max}$ = 2 : a saddle point where isocontours cross, and where the inner and outer allowed zones interconnect when the kinetic energy of particles is larger than 1/32 - in Dragt's dimensionless length unit - . When the energy of an interplanetary particle is larger than 1/32 it can override the magnetic potential barrier, and penetrate deep into the geomagnetic field between cusp-like dashed lines. These distributions are drawn for charged particle characterized by
$\varepsilon = sign(-p_o/Z) = +1$ [Dragt, 1965].

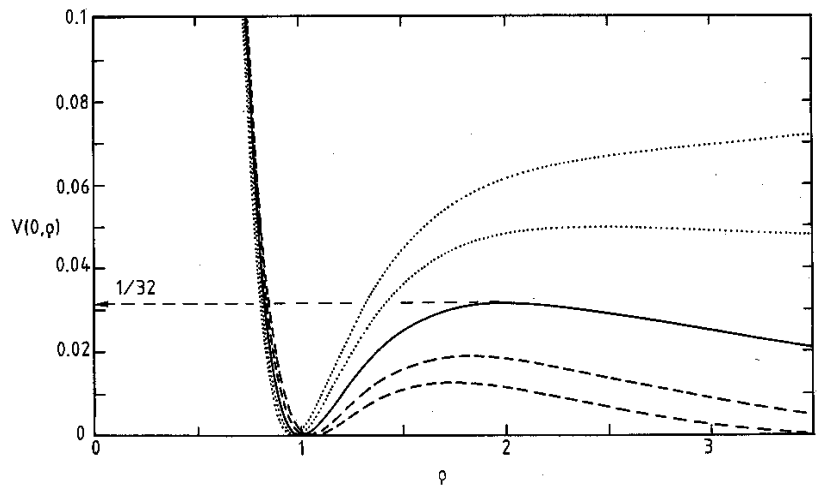

Figure 2. *Equatorial cross-section of the extended Störmer potential, V(ρ,0), as a function of the equatorial distance ρ.* When the value of the IMF is equal to zero the maximum value of Störmer's potential is equal to *V = 1/32 = 0.03125*. For *F ≠ 0* this maximum value and its position vary as described in the text. The two lower curves lines correspond to southward IMF: *b = -0.03* and *-0.05*. The two upper ones correspond to northward IMF: *b = +0.03* and *+0.05* [Lemaire, 2003].

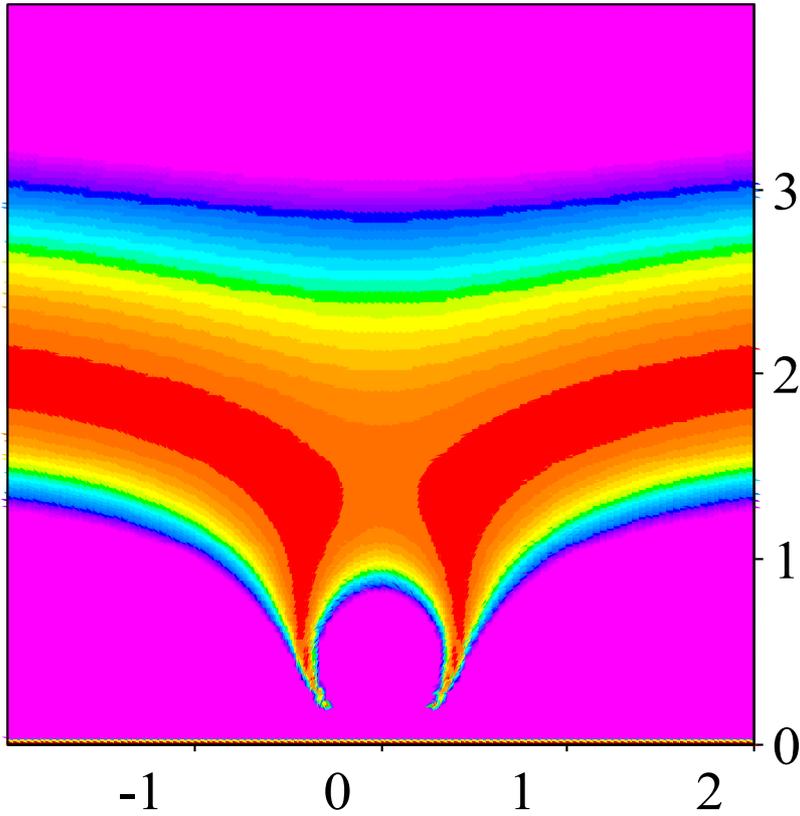

Figure 3. *Extended Störmer potential, V(ρ, z) for a large (negative) southward IMF : b = -0.21, ( i.e.: F = - 13.4 nT when $r_u$ = 10 $R_E$ ).* The neutral or X-line is then at the equatorial distance $ρ_X = 1/(–2 b)^{1/3} = 1.33$ in units $r_u$. In this case the Thalweg is formed of two semi-infinite branches extending to infinity along two interconnected magnetic field lines. Along these open Thalwegs segments (not drawn) interplanetary particles of all energies can enter and penetrate deep into the geomagnetic field [Lemaire, 2003].